\documentclass[12pt]{article}
\usepackage{amsmath, amssymb}
\usepackage{graphicx}
\usepackage{subcaption}
\usepackage{epsfig}
\usepackage{color}
\usepackage{amsthm}
\usepackage{amsopn}
\usepackage{url}
\usepackage[numbers]{natbib}
\usepackage{hyperref}
\usepackage{cleveref}

\setcitestyle{numbers}

\usepackage[top=2cm, bottom=2cm, left=2.5cm, right=2.5cm]{geometry}
\title{A General Framework for Leggett--Garg Violation in Neutrino Oscillations and Matter Effects}

\author{
	M. M. Ettefaghi\thanks{Corresponding author: \url{mettefaghi@qom.ac.ir}}\, and H. Hajipour \\
	\small Department of Physics, University of Qom, Qom 371614-6611, Islamic Republic of Iran
}

\begin{document}
	
	\maketitle
	
	\begin{abstract}
		The Leggett--Garg inequality (LGI) constrains temporal correlations under the assumptions of macrorealism and noninvasive measurability.
		We study the four-time quantity \(K_4\) for two-flavor neutrino oscillations using a pseudo-density matrix (PDM) under static conditions, instead of non-invasive measurement. With this approach, it becomes possible to examine a Horodecki-type criterion for the temporal Pauli-correlation tensor. The Horodecki criterion had been proven for CHSH inequality (R. Horodecki et al. Phys. Lett. \textbf{A200}, 340), For any unitary qubit evolution, the corresponding scalar is \(M_R=2\).
		Although, this result characterizes the correlation tensor and the optimized temporal construction, it does not imply that a fixed flavor observable violates the LGI at every neutrino energy or baseline. We therefore evaluate \(K_4\) explicitly for flavor measurements in matter of constant density. The numerical examples illustrate how matter effects modify the temporal correlations, with a pronounced matter--vacuum separation near the MSW resonance for the chosen parameters and generally smaller differences in the selected off-resonance cases. This finding motivates further investigation of matter-induced temporal correlations in atmospheric neutrinos and their potential relevance to future studies with long-baseline and atmospheric-neutrino experiments, including DUNE and Hyper-Kamiokande.
		
	\end{abstract}
	
	\section{Introduction}
	Bell's inequality is a fundamental tool for distinguishing quantum mechanical phenomena that are in conflict with the postulate of local realism \cite{Bell1964}. Although local realism is a natural postulate in classical theories, it is challenged in quantum mechanics. Bell's inequality focuses on spatial correlations between entangled particles. The Leggett-Garg inequality (LGI), sometimes called the time-like version of Bell's inequality, is based on temporal correlations within a single system to examine macroscopic realism (macrorealism) and noninvasive measurability \cite{Leggett1985}. Specifically, the LGI investigates whether the behavior of a system over time can be described by a classical trajectory, assuming that the system exists in a definite state at all times and that measurements do not disturb its evolution.
	
	In particle physics, numerous phenomena can be utilized to explore the Bell and Leggett-Garg inequalities \cite{Go2004,Bertlmann2004,Bramon2005,fabbrichesi2024,barr2024quantum}. Notably, since neutrino survival probability is less than unity due to oscillations during propagation, it has been demonstrated that neutrino oscillation experiments imply a violation of Bell's inequality \cite{Alok2016}. This result was established by a theorem developed by the Horodecki family in Ref. \cite{Horodecki1995}, which introduced a quantitative criterion---hereafter referred to as the Horodecki parameter---derived from the system’s density operator to characterize violations of the CHSH (Clauser–Horne–Shimony–Holt) form of Bell’s inequality \cite{Clauser1978}.
	
	Beyond testing the bounds of local realism, neutrino oscillations serve for investigating fundamental quantum correlations. The inherent superposition and mixing of mass eigenstates naturally give rise to flavor entanglement, quantum coherence, and non-classical correlations \cite{ettefaghi2020quantum,blasone2021wave,bachain2026dephasing,alok2016quantum}. In recent years, there has been growing interest in characterizing these quantum features using information-theoretic measures. For instance, several studies have analyzed the entanglement, quantum discord, and coherence generated during neutrino flavor evolution in both two-flavor and three-flavor frameworks \cite{ettefaghi2023quantum, banerjee2025analysis, banerjee2026concurrence,jha2026quantum,benerjee2026wave,yadav2025impact,li2022geuine,dixit2024quantum}. These works explore how mass-mixing parameters, CP-violating phases, and propagation through matter influence the generation and degradation of quantum correlations \cite{bouri2025probing, dixit2018quantum}.
	Beyond characterizing the generation and degradation of quantum correlations, such measures have also been shown to have practical implications for precision neutrino physics; in particular, the local minima of flavor entanglement in three-flavor long-baseline oscillations have been proposed as optimal regions for extracting oscillation parameters, offering improved constraints on $\sin^2\theta_{23}$, $\delta_{\rm CP}$, and $\Delta m^2_{31}$ and affecting sensitivities to leptonic CP violation, the $\theta_{23}$ octant, and the mass ordering \cite{dixit2026quantum}. Complementing these studies on spatial and informational correlations, the temporal correlations within a single neutrino system make this phenomenon highly suitable for studying the violation of the LGI.
	
	Neutrino oscillation arises due to the quantum mechanical superposition of mass eigenstates, and consequently, the survival probability for a given neutrino flavor becomes time-dependent. Therefore, this phenomenon is highly suitable for studying the violation of the LGI. For example, in the Refs. \cite{Formaggio2016} and \cite{Fu2017}, the violation of the LGI has been shown using the MINOS and Daya-Bay experiments, respectively. In addition, various other papers have addressed LGI violations in neutrino oscillation theory under different conditions \cite{Richter2017,Gangopadhyay2017,Naikoo2019}. For instance, assuming a dissipative interaction exists between matter and neutrinos, one can demonstrate that the $K_3$ function, introduced in the context of the Leggett-Garg inequality violation (the subscript 3 indicates that the measurement was performed at 3 different times), takes on different values depending on whether the neutrino is a Dirac or Majorana particle \cite{Richter2017}. Using weak field approximations, it has been shown that gravitational effects reduce the maximum value of $K_3$ within certain energy intervals, leading to the absence of LGI violation \cite{Ettefaghi2022,wu2026interpreting}. More recently, the characterization of macrorealism in neutrino oscillations has been extended beyond the standard Leggett–Garg inequalities to stronger necessary and sufficient conditions, namely the no-signaling-in-time and arrow-of-time conditions, which have been examined in both two- and three-flavor frameworks, including the CP-violating phase and the wave packet approach \cite{blasone2023no,Blasone2026quantum}. A recent reappraisal by Groth et al. emphasizes that a statistically meaningful experimental LGI test requires explicit macrorealistic background models: apparent violations in reconstructed correlators are not, by themselves, sufficient to establish statistically significant rejection of macrorealism \cite{groth2025neutrino}. This methodological point concerns the statistical interpretation of experimental data and is distinct from calculating ideal quantum correlators for a specified Hamiltonian.
	
	A central difficulty in applying the LGI to neutrino oscillations is the assumption of noninvasive measurability (NIM), which requires that a measurement reveal the pre-existing value of an observable without disturbing the subsequent evolution of the system. Flavor measurements generally alter the neutrino's quantum state, making the direct implementation of NIM challenging. Moreover, successive flavor measurements on the same neutrino are experimentally impractical.  To address these difficulties, many studies of LGI in neutrino oscillations adopt the alternative assumption of stationarity \cite{Formaggio2016,Fu2017,Richter2017,Gangopadhyay2017,Naikoo2019}.    Under this assumption, temporal correlators depend only on the time difference between measurements \cite{Emary2013},
	\[
	C(t_i,t_j)=C(0,t_j-t_i),\qquad t_j>t_i.
	\]
	Consequently, the required correlators can be reconstructed from flavor-transition or survival probabilities measured using an ensemble of similarly prepared neutrinos at different propagation baselines, rather than by performing successive measurements on the same neutrino. This approach makes LGI-type tests experimentally accessible, but stationarity is an additional assumption, not a general proof of noninvasiveness. As discussed in Ref.~\cite{Emary2013}, the resulting test is therefore restricted to macrorealistic models compatible with the adopted stationarity condition, rather than the full class of models constrained by the original LGI under NIM.
	The interpretation of such tests also requires care at the statistical level. Groth et al. have shown that assessing the significance of an apparent LGI violation in neutrino data requires explicit macrorealistic background models and a treatment of statistical fluctuations; a violation reconstructed from measured correlators alone does not establish statistically significant rejection of macrorealism \cite{groth2025neutrino}. This issue is distinct from calculating ideal quantum correlators for a specified Hamiltonian, as done in the present theoretical analysis.
	
	On the other hand, the propagation of neutrinos in a material medium leads to an additional potential energy $V_e$ for electron neutrinos. 
	This potential arises from elastic forward scattering via electroweak interactions, preserving the coherence of neutrino waves and maintaining interference effects. When matter density varies adiabatically, a resonance phenomenon can occur, allowing almost complete flavor transition even for small mixing angles. This effect, known as the MSW effect, was first described by \cite{Wolfenstein1978} and later expanded by \cite{Mikheev1985}, explaining solar neutrino oscillations. However, at very high matter densities, the transition probability of electron neutrinos can diminish \cite{Ravari2022}.

	In this paper, we use the pseudo-density matrix (PDM) formalism \cite{Fitzsimons2015} to investigate four-time temporal correlations in two-flavor neutrino oscillations and examine how matter effects influence the Leggett--Garg parameter $K_4$, whose subscript denotes the four measurement times entering the temporal correlation function. Following the approach commonly adopted in neutrino LGI studies, we use the stationarity assumption to construct the required temporal correlators from oscillation probabilities at different baselines. We then adopt the Horodecki criterion, originally introduced for the CHSH inequality \cite{Horodecki1995}, and extend it to the temporal domain via the PDM formalism \cite{jalal2026quantum}, in order to establish a necessary and sufficient condition for LGI violation. In this framework, the optimized LGI parameter is expressed as $K_4^{\max}=2\sqrt{M_R}$, where $M_R$ is constructed from the temporal Pauli-correlation tensor; hence, $M_R>1$ is the necessary and sufficient condition for violation within the specified optimization. We prove that $M_R=2$ for any unitary qubit evolution, showing that the optimized temporal construction permits maximal LGI violation in this setting. This result, however, does not imply that the LGI is violated for a fixed flavor observable at arbitrary energies, baselines, or matter densities. The flavor-based $K_4$ must be evaluated separately for the physical parameters of interest. In particular, the no-mixing limit illustrates this distinction: although $M_R=2$, the flavor-based $K_4$ reaches but does not exceed the classical bound. Therefore, it is natural to examine how matter-induced changes in the effective mixing angle and oscillation phase affect $K_4$, including the behavior near the MSW resonance.
	
	Therefore, we examine how the matter-modified mixing angle and oscillation phase alter the explicit flavor-based $K_4$. To construct the four correlators from baseline-dependent data, we use the stationarity assumption discussed in the LGI literature \cite{Emary2013,Formaggio2016}. The baseline pattern $L_a=L$, $L_b=3L$, $L_c=5L$, and $L_d=9L$ is used as a convenient phase-additive configuration for the constant-density calculation, not as a universally optimal experimental prescription. The numerical examples compare vacuum and matter predictions at fixed parameter choices. They show that the matter--vacuum difference can become pronounced near the MSW resonance, while the selected off-resonance examples generally exhibit a smaller difference. We do not model realistic Earth trajectories, detector response, statistical fluctuations, or event counts here; those aspects are left for a future phenomenological analysis.
	
	The paper is organized as follows: In Section~\ref{sec:pseudo}, we introduce the pseudo-density matrix formalism and its application to two-flavor neutrino oscillations. Section~\ref{sec:lgi} presents the derivation of two-time correlation functions and the construction of the LGI parameter $K_4$. In Section~\ref{sec:calculations}, we establish the generalized Horodecki criterion for the neutrino system and present a numerical study of $K_4$ in both vacuum and matter, analyzing the MSW resonance enhancement. Finally, in Section~\ref{sec:conclusion}, we discuss the implications of our results, briefly outline future prospects for experimental tests, and conclude.
	
	\section{PDM Formalism for Neutrino Oscillations}\label{sec:pseudo}
	To calculate temporal correlations, an extension of the standard density matrix formalism, which commonly used to describe statistical mixtures of quantum states and their evolution, was developed in the form of the PDM formalism. A PDM is defined as a Hermitian, unit-trace matrix whose marginals reduce to ordinary density matrices. This approach enables the analysis of quantum systems at multiple times without assuming noninvasive measurability, a key requirement for testing the LGI \cite{Fitzsimons2015}. Consequently, the PDM framework provides a natural and rigorous tool for investigating time-dependent quantum phenomena, including processes relevant to neutrino oscillations and propagation through matter.
	
	\subsection{Definition and Construction}
	
	Let's consider a system on which several Von Neumann measurements have been performed at times $t_{1}, t_{2}, \ldots, t_{n}$. Each Von Neumann measurement can be represented by single-Qubit Pauli matrices. We consider a sequence of measurements $\left\lbrace \sigma_{i_{j}}\right\rbrace _{j=1}^{n}=\left\lbrace \sigma_{i_{1}}\cdots \sigma_{i_{n}}\right\rbrace$ performed on the system at the corresponding times.  After the first measurement, the system collapses to one of $\sigma_{i_{1}}$ eigenstates and then undergoes a time evolution until the second measurement. At this stage, the system again collapses to one of $\sigma_{i_{2}}$ eigenstates  and undergoes a time evolution until the next measurement. This process continues until the n-th measurement step. The PDM $R$ is defined as \cite{Fitzsimons2015}:
	\begin{equation}
		R = \frac{1}{2^{n}} \sum_{i_{1} = 0}^{3} \cdots \sum_{i_{n} = 0}^{3} 
		\left\langle \{\sigma_{i_{j}}\}_{j=1}^{n} \right\rangle 
		\left( \bigotimes_{j=1}^{n} \sigma_{i_{j}} \right),
	\end{equation}
	where $\bigotimes$ denotes the tensor product, and $\left\langle \{\sigma_{i_{j}}\}_{j=1}^{n} \right\rangle$ is the expectation value of the mentioned sequence of measurements.
	
	The main difference between a PDM $R$ and a conventional density matrix is that $R$ is not necessarily positive semi-definite. For a single-time system, the PDM reduces to the standard density matrix $\rho$, while for multiple times, $R$ generalizes $\rho$ by incorporating temporal correlations. For example, for two times $t_{1}$ and $t_{2}$, the pseudo-density matrix $R$ can be written as:
	\begin{equation}
		R = \frac{1}{4} \sum_{i,j} \left\langle \{\sigma_{i}(t_{1}) \sigma_{j}(t_{2})\} \right\rangle 
		\left( \sigma_{i} \otimes \sigma_{j} \right).\label{R12}
	\end{equation}
	
	In Ref. \cite{jalal2026quantum}, it is shown that for a single-qubit system with density matrix $\rho$ and unitary time evolution $U$ under arbitrary measurements at times $t_{1}$ and $t_{2}$, we have:
	\begin{equation}
		\left\langle \{\sigma_{i}(t_{1}) \sigma_{j}(t_{2})\} \right\rangle = 
		\frac{1}{2} \text{Tr} \left[ (\sigma_{i}\rho + \rho\sigma_{i}) U^{\dagger}\sigma_{j}U \right].\label{R12coeffecient}
	\end{equation}
	In cases of non-unitary time evolution, such as in open quantum systems, the density matrix formalism must incorporate decoherence effects.
	
	\subsection{PDM for Two-Flavor Neutrino oscillation Schema}\label{sec:twoflavor}
	Neutrino oscillation is a quantum mechanical phenomenon in which the flavor of a neutrino can change as it propagates. Although a neutrino is produced in a definite flavor state, this state is actually a coherent superposition of different mass eigenstates because neutrino masses are extremely small in comparison to the intrinsic energy and momentum uncertainties. As these mass eigenstates evolve with slightly different phases over time, the flavor composition of the neutrino changes, giving a non-zero probability of observing a different flavor at detection.
	
	In the two-flavor approximation, which captures the essential physics while avoiding unnecessary 
	complexity, a neutrino created  in a definite flavor state (e.g., $\nu_e$ or $\nu_\mu$) is expressed as a superposition of two mass eigenstates $\nu_1$ and $\nu_2$. The phase difference acquired during propagation leads to oscillations between flavors. This simple framework yields the standard oscillation probability, characterized by two key parameters: the mixing angle $\theta$ and the squared mass difference $\Delta m^2$. Despite its simplicity, the two-flavor scheme provides an excellent first approximation for many experimental situations and serves as the foundation for understanding the richer three-flavor oscillation picture.
	
	The effective Hamiltonian for the two-flavor schema (with $\hbar = c = 1$) is:
	\begin{equation}
		\mathcal{H} = \left( p + \frac{m_{1}^{2} + m_{2}^{2}}{4p} + \frac{V_{C}}{2} + V_{N} \right) 1 + 
		\frac{1}{2} \begin{pmatrix}
			V_{C} - \omega\cos{2\theta} & \omega\sin{2\theta} \\
			\omega\sin{2\theta} & \omega\cos{2\theta} - V_{C}
		\end{pmatrix}\equiv r_{0} 1 + \frac{\vec{r} \cdot \vec{\sigma}}{2}
		\label{eq:Ham}
	\end{equation}
	Here, \( \theta \) denotes the vacuum mixing angle, while \( m_1 \) and \( m_2 \) are the masses of the two neutrino mass eigenstates. The oscillation frequency is defined as $\omega \equiv\frac{\Delta m^2 }{2p}$, where $p \approx E$ is the neutrino momentum. The squared mass splitting that governs 
	the atmospheric neutrino oscillation, first measured by the Kamiokande experiment, is \( \Delta m^2_{32} \approx 2.5 \times 10^{-3} \, \text{eV}^2 \), which is approximately 30 times larger than the solar splitting measured by the SNO experiment, \( \Delta m^2_{21} \approx 7.4 \times 10^{-5} \, \text{eV}^2 \). The typical value of the mixing angles are \( \theta_{12} \approx 33.46^\circ \) and \( \theta_{13} \approx 8.56^\circ \). The term $V_{C(N)}=\sqrt{2}G_Fn_{e(n)}$
	is the charged (neutral) current potential due to coherent
	forward scattering of neutrinos with electrons (neutrons)
	in matter, and $G_F$ is the Fermi coupling constant.
	
	For a neutrino with energy $(E_{a})$, the flavor state evolves in time under a unitary operator $\mathcal{U}$, which is determined by the oscillation part of the Hamiltonian, \begin{equation}
		\mathcal{H}_{osc}=\frac{\vec{r} \cdot \vec{\sigma}}{2}.
	\end{equation}
	The corresponding time-evolution operator between $t_{i}$ and $t_{j}$ is:
	\begin{equation}
		\mathcal{U}(\psi_{a;ij}) = \exp\left[ -i \int_{t_{i}}^{t_{j}} \mathcal{H}_{osc}(\omega_{a}) \, dt \right] \approx 
		\cos(\psi_{a;ij}) 1- i \sin(\psi_{a;ij}) (\hat{r}(\omega_{a}) \cdot \vec{\sigma})
	\end{equation}
	where
	\begin{equation}\psi_{a;ij} = |\vec{r}(\omega_{a})| \frac{t_{j} - t_{i}}{2}, \label{psiij}\end{equation} is the phase acquired by a neutrino with energy $E_{a}$.
	
	For the initial state $\rho = |\nu_{e}\rangle \langle \nu_{e}|$, using (\ref{R12}) and (\ref{R12coeffecient}), one can obtain the PDM for two measurement at times $t_i$ and $t_j$ as follows:
	\begin{equation}
		R(t_i,t_j) = \begin{pmatrix}
			|Z_{ij}|^2 & -\sin(2\theta) S_{ij} Z_{ij} & - \frac{1}{2} \sin(2\theta) S_{ij} Z_{ij}^* & \frac{1}{2}  \sin^{2}(2\theta) S_{ij}^{2}\\
			- \sin(2\theta) S_{ij} Z_{ij}^* & \sin(2\theta)^{2} S_{ij}^{2} & -\frac{1}{2} {Z_{ij}^*}^{2}  & \frac{1}{2} \sin(2\theta) S_{ij} Z_{ij}^* \\
			- \frac{1}{2} \sin(2\theta) S_{ij} Z_{ij} & -\frac{1}{2} Z_{ij}^{2} & 0 & 0 \\
			\frac{1}{2} \sin^{2}(2\theta) S_{ij}^{2} & \frac{1}{2} \sin(2\theta) S_{ij} Z_{ij} & 0 & 0
		\end{pmatrix},\label{Rij}
	\end{equation}
	where $S_{ij} \equiv \sin(\psi_{a;ij})$, $C_{ij} \equiv \cos(\psi_{a;ij})$ and
	$Z_{ij}\equiv\cos(2\theta) S_{ij}-i C_{ij}$.
	The diagonal terms correspond to survival $P_{ee}(t)$ and transition $P_{e\mu}(t)$ probabilities.
	
	\section{LGI in the Framework of PDM Formalism}\label{sec:lgi}
	In contrast to the Bell inequality which probes spatial correlations, the LGI constrains, under certain fundamental assumptions, the correlations between measurements of a quantity $Q$ at different times. Unlike the conventional density matrix, which only describes spatial correlations at a fixed time, the PDM generalizes the formalism to incorporate multi-time statistics without assuming noninvasive measurements. This makes it especially well-suited for analyzing the conditions under which the LGI is violated and for quantifying temporal quantum correlations in systems with time-dependent dynamics, such as neutrino oscillations or open quantum systems. Thus the PDM formalism provides a natural framework for studying such temporal correlations.  
	
	\subsection{LGIs}\label{sec:intro}
	As mentioned, the LGIs were originally formulated to test the compatibility of quantum mechanics with two classical assumptions: macroscopic realism (the idea that a system possesses definite properties independent of observation) and noninvasive measurability (the possibility of determining these properties without disturbing the system). Therefore, a violation of these inequalities serves as an evidence of non-classical behavior, in particular quantum coherence persisting across different times. This is closely analogous to the violation of Bell’s inequality, which signifies entanglement as a form of spatial quantum correlation. 
	
	Let's consider a quantum system on which some dichotomic observable operators denoted by \( \hat{Q} \), whose realization outcomes take values \( \pm1 \), can be measured.  The corresponding two-time correlation function is defined as follows: 
	\begin{equation}\label{1}
		C_{ij}\equiv \langle \hat{Q}_{i}\hat{Q}_{j}\rangle =\frac{1}{2}\langle\phi|\{\hat{Q}_{i},\hat{Q}_{j}\}|\phi\rangle. \hspace{1 cm} \hat{Q}_{i(j)}\equiv \hat{Q}(t_{i(j)})
	\end{equation}
	The classical counterpart of this definition is written as follows:
	\begin{equation}\label{11}
		C_{ij}=\sum_{Q_{i},Q_{j}=\pm1}^{ } Q_{i}Q_{j}P_{ij}(Q_{i},Q_{j}), 
	\end{equation}
	where, $P_{ij}(Q_{i},Q_{j})$ represents the joint probability of obtaining \( Q_i \) and \( Q_j \) as measurement outcomes at times \( t_i \) and \( t_j \), respectively. 
	
	The Leggett-Garg parameter for $n$ measurements is defined as:
	
	\begin{equation}
		K_{n} \equiv \sum_{i = 1}^{n - 1}{C_{i,i + 1} - C_{n,1}},\label{kn}
	\end{equation}
	
	where $C_{i,j}$ denotes the two-time correlation function. According to the LG inequality, for classical systems, this parameter must satisfy \cite{Leggett2008}:
	
	\begin{equation}
		-(n-2) \leq K_{n} \leq n - 2 \quad \text{for } n \geq 3. \label{LGIsn}
	\end{equation}
	To study the behavior of neutrino oscillations, we need to derive the Leggett–Garg parameters from two-time correlation functions.
	
	\subsection{Derivation of Two-Time Correlation Functions}
	\label{sec:correlation}
	The PDM formalism offers a natural framework for analyzing temporal correlations in quantum systems.
	Accordingly, temporal correlations like Equation \eqref{1} can be expressed as follows:
	\begin{equation}\label{22} 
		C_{ij}=\text{Tr}\bigg(R(t_i,t_j) \hat{Q}_{i} \otimes \hat{Q}_{j} \bigg). 
	\end{equation}
	Here, \( R(t_i,t_j) \) represents the PDM at the initial time \( t_i \) and the subsequent time \( t_j \).
	
	For a two-flavor neutrino system, we consider the observable $\widehat{Q} \equiv \sigma_{z}$ (the third component of Pauli matrices) representing the neutrino flavor. This operator has eigenvalues +1 and -1, corresponding to electron and muon flavor states, respectively.
	So, using the PDM given in Eq. (\ref{Rij}), the correlation function is obtained by choosing $Q_i=Q_j\equiv\sigma_z$ as follows: 
	
	\begin{equation}
		C\left( t_{i},t_{j} \right) = 1 - 2 \sin^2(\psi_{a;i,j}) \sin^2 (2\theta).\label{ncij}
	\end{equation}
	As expected, this result agrees with that obtained in Ref. \cite{Formaggio2016}. As seen from the definition of $\psi_{a;ij}$ in Eq. (\ref{psiij}), the notable feature of this result is that the temporal correlation depends solely on the time difference $\tau = t_{j} - t_{i}$.
	This phase encodes the dynamics of the quantum state and generally incorporates both vacuum oscillations and matter effects. Therefore, in neutrino oscillation, temporal quantum correlations can then be probed by studying how they vary with different spacetime intervals between measurements.
	
	In the following, we consider the LGI for $K_4$, which from Eq. (\ref{LGIsn}) can be written as
	\begin{equation}
		-2 \leq K_4\leq 2.\label{k4}
	\end{equation}
	This inequality is analogous to the CHSH version of Bell’s inequality \cite{Clauser1978}.
	According to Eq. (\ref{kn}), $K_4$ is given by
	\begin{equation}
		K_{4} = C_{1,2} + C_{2,3} + C_{3,4} - C_{1,4},
	\end{equation}
	where $t_{1} < t_{2} < t_{3} < t_{4}$. 
	
	\section{Calculation of LGIs in Neutrino Oscillations}\label{sec:calculations}
	In order to examine the quantum nature of neutrino oscillations, we now turn to the explicit calculation of the LGIs in this framework. Using the two-time correlation functions defined for flavor measurements given in Eq. (\ref{ncij}), we construct the LGI parameters and evaluate their behavior under neutrino time evolution. This allows us to identify conditions under which the LGIs are violated, thereby revealing signatures of quantum coherence in neutrino oscillations.
	
	\subsection{LGI Violation and Horodecki Theorem}\label{sec:horodecki}
	Although the violation of Bell’s inequality is well understood for pure states, its extension to mixed states remained ambiguous until the Horodecki family established a powerful theorem in 1995 \cite{Horodecki1995}. This result provides a necessary and sufficient criterion for detecting violations of the CHSH version of Bell’s inequality in arbitrary two-qubit mixed states, expressed directly in terms of the density matrix of the system. The strength of this approach lies in reducing the problem of testing Bell inequalities from a dependence on complicated measurement settings to an intrinsic property of the quantum state itself. A corresponding criterion can be used for the temporal PDM construction considered here, where the tensor-product operator basis labels two measurement times rather than two independently prepared physical qubits \cite{jalal2026quantum}. 
	
	Generally, the PDM of a two-time two-state system is similar to a $4\times4$ two-Qubit system. Therefore, it can be written based on the Pauli matrices of two qubits as follows:
	\begin{equation} \label{29} 
		R_{ij}=\frac{1}{4}\bigg(\mathbb{I}_{i}\otimes \mathbb{I}_{j}+\bf{r}.\bf{\sigma}_{i}\otimes \mathbb{I}_{j}+\mathbb{I}_{i}\otimes \bf{s}.\bf{\sigma}_{j}+\sum_{m,n=1}^{3}t_{mn}\sigma_{i_{m}}\otimes\sigma_{j_{n}}\bigg).
	\end{equation}
	Here, the indices $i$ and $j$ label the measurement stages, while $\mathbf r$ and $\mathbf s$ are real three-vectors. The temporal Pauli-correlation tensor \( T_R \) is constructed by \( t_{mn}=\text{Tr}(R_{ij}\sigma_{i_{m}}\otimes\sigma_{j_{n}}) \) coefficients. We define the scalar parameter \( M_R \) as follows:
	\begin{equation} \label{30} 
		M_{R}=u_{k}+u_{l}
	\end{equation}
	Here, \( u_k \) and \( u_l \) represent two largest eigenvalues of matrix $U=T_{R}^{\dagger}T_{R}$. Similar to the approach outlined in reference \cite{Horodecki1995}, for the temporal PDM, optimizing over the relevant dichotomic measurement directions gives the maximum value \(K_4^{\max}=2\sqrt{M_R}\) \cite{jalal2026quantum}. Thus $M_R>1$ is necessary and sufficient for the existence of measurement settings for which $|K_4|$ exceeds the classical upper bound 2.

	Let us consider a qubit with an arbitrary initial density matrix $\rho=(I+\mathbf r\cdot\boldsymbol{\sigma})/2$ and unitary evolution $U$ between the two times. Using Eqs. (\ref{R12}) and (\ref{R12coeffecient}), one can write the temporal Pauli-correlation tensor components as follows:
	\begin{equation}
		(T_R)_{\alpha\beta}=\frac12\operatorname{Tr}\!\left[\{\sigma_\alpha,\rho\}U^\dagger\sigma_\beta U\right],\qquad \alpha,\beta=1,2,3.
		\label{eq:TRunitary}
	\end{equation}
	Unitary conjugation rotates the Pauli operators, so there is a real orthogonal matrix $O$ such that
	\begin{equation}
		U^\dagger\sigma_\beta U=\sum_{\gamma=1}^{3}O_{\beta\gamma}\sigma_\gamma,
		\qquad O^{\mathsf T}O=I_3.
		\label{eq:paulirotation}
	\end{equation}
	Using the Bloch form of $\rho$ and the Pauli anticommutation relation gives
	\begin{equation}
		\{\sigma_\alpha,\rho\}=\sigma_\alpha+r_\alpha I,
		\qquad \operatorname{Tr}(\sigma_\alpha\sigma_\gamma)=2\delta_{\alpha\gamma},
	\end{equation}
	and hence
	\begin{align}
		(T_R)_{\alpha\beta}
		&=\frac12\sum_\gamma O_{\beta\gamma}
		\operatorname{Tr}\!\left[(\sigma_\alpha+r_\alpha I)\sigma_\gamma\right]\nonumber\\
		&=O_{\beta\alpha}.
	\end{align}
	Therefore $T_R=O^{\mathsf T}$ and
	\begin{equation}
		T_R^{\mathsf T}T_R=I_3.
	\end{equation}
	All three eigenvalues of $S_R=T_R^{\mathsf T}T_R$ are consequently equal to unity, $u_1=u_2=u_3=1$, and the scalar defined in Eq.~\eqref{30} is
	\begin{equation}
		M_R=u_1+u_2=2.
		\label{eq:MRunitary}
	\end{equation}	
	This result shows that, for an arbitrary initial qubit state and an arbitrary Hamiltonian governing unitary time evolution, the LGI is maximally violated, \(K_4^{\max}=2\sqrt{2}\), when the optimized directions are chosen for measuring the relevant dichotomic operators.

	For the neutrino oscillations described in Section~\ref{sec:pseudo}, according to the result established in this section, the unitary evolution leads to the LGI violation provided that we have optimized temporal construction. However, this result does not guarantee violation for an arbitrary choice of the observable $Q$. In the following, we consider flavor measurements by setting $Q=\sigma_3$ at all times. This fixed choice is not necessarily optimal for maximizing $K_4$. Therefore, whether the LGI is violated, and how closely $K_4$ approaches its quantum maximum, must be determined from the explicit flavor-based calculation presented below.

	\subsection{Numerical Results}\label{sec:numerical}
	
	One of the main practical obstacles to testing the Leggett–Garg inequality is that Eq.~\eqref{kn} requires measurements at more than two different times, whereas a typical neutrino oscillation experiment provides data at only two locations: a near detector and a far detector. As discussed in Ref.~\cite{Formaggio2016}, this obstacle can be circumvented by replacing the noninvasive-measurability assumption with a \emph{stationarity assumption}, under which the two-time correlation function depends only on the time separation between measurements rather than on their absolute values. Under this assumption, the correlation functions at different time separations can be reconstructed from an ensemble of identically prepared neutrinos observed at different baselines, rather than from repeated noninvasive measurements performed on a single system. As emphasized by Groth et al. \cite{groth2025neutrino}, predicting an LGI violation theoretically is distinct from establishing its statistical significance in an experiment. Although a theoretical value satisfying $|K_4|>2$ indicates a violation of the classical bound, an experimental claim requires an appropriate statistical analysis that accounts for fluctuations and relevant macrorealistic models. The present work focuses on idealized quantum predictions and does not claim an experimental rejection of macrorealism.
	
	To illustrate the formalism, we consider two representative parameter sets corresponding to the conventional solar and atmospheric mass-squared splitting scales in the two-flavor framework. Although, in long-baseline accelerator experiments, the source-to-detector distance is fixed, we draw upon the point that atmospheric neutrinos reach the detector over a continuous range of baselines corresponding to their different zenith angles. This feature makes it possible to construct the several two-time correlation functions needed to evaluate the LGI parameter $K_4$.
	Moreover, the two-time correlation function \(C_{ij}\), as given by Eqs.~\eqref{psiij} and \eqref{ncij}, depends only on the time interval \(\tau \equiv t_j-t_i\), consistent with the stationarity assumption introduced above. A further simplification follows from the \emph{sum rule}, which relates correlation functions evaluated over different time intervals and expresses the additivity of the accumulated oscillation phase:
	\begin{equation}
		\sum_{i=1}^{n-1} \psi_{a;\,i\,i+1} = \psi_{a;\,1\,n}. \label{eq:sumrule}
	\end{equation}
	As shown in Eq.~\eqref{eq:Ham}, the effective Hamiltonian in matter contains an additional potential term arising from the coherent forward scattering of neutrinos on electrons. This potential shifts both the mixing angle and the oscillation frequency away from their vacuum values. Diagonalizing the Hamiltonian of Eq.~\eqref{eq:Ham} gives the effective mixing angle in matter,
	\begin{equation}
		\tan 2\theta_m = \frac{\Delta m^2 \sin 2\theta}{\Delta m^2 \cos 2\theta - 2EV}, \label{eq:tan2thetam}
	\end{equation}
	together with the effective mass-squared splitting,
	\begin{equation}
		\Delta m_m^2 = \Delta m^2 \sqrt{\left( \cos 2\theta - \frac{2EV}{\Delta m^2} \right)^2 + \sin^2 2\theta}. \label{eq:dm2m}
	\end{equation}
	Both effective parameters are energy dependent and, together, fully determine the oscillation dynamics inside matter. Here $V = \sqrt{2}\, G_F n_e$, where $n_e = N_A Y_e \rho$ is the electron number density along the neutrino trajectory, $N_A$ is Avogadro's number, and $Y_e$ is the electron fraction, and $\rho$  is the density of the medium.
	The most striking consequence of these matter effects is the MSW resonance, at which flavor conversion is strongly enhanced. The resonance condition is reached when
	\begin{equation}
		2EV = \Delta m^2 \cos 2\theta, \label{eq:resonancecond}
	\end{equation}
	which corresponds to the resonance energy
	\begin{equation}
		E_{\text{res}} = \frac{\Delta m^2 \cos 2\theta}{2\sqrt{2}\, G_F N_A Y_e \rho}. \label{eq:Eres}
	\end{equation}
	At resonance, the effective mixing angle attains its maximal value,
	\begin{equation}
		\theta_m = \frac{\pi}{4}, \label{eq:thetamax}
	\end{equation}
	irrespective of the vacuum mixing angle $\theta$. Thus, the amplitude of flavor conversion is enhanced in the constant-density two-flavor model. This can produce a pronounced change in the temporal correlators entering $K_4$. However, the value of $K_4$ also depends on the accumulated oscillation phases and the baseline configuration, so maximal effective mixing alone does not guarantee an LGI violation or a maximum of $K_4$ in every setup. The resonance region may therefore exhibit a pronounced departure from the classical LGI bound, depending on the accumulated phases and the chosen baseline configuration.

	We now specify the physical inputs used in the numerical analysis of matter
	effects on the violation of the LGI. We retain two parameter sectors: the solar mass-splitting sector, described by $\Delta m_{21}^2 = 7.4 \times 10^{-5}\ \mathrm{eV}^2$ and$\theta_{12} \simeq 33.46^\circ$
	($\sin^2 2\theta_{12} = 0.86$), and the atmospheric-scale sector, in which the electron-flavor matter resonance is controlled by $\theta_{13} \simeq 8.56^\circ$
	($\sin^2 2\theta_{13} = 0.087$) and an atmospheric splitting approximated here by $\Delta m^2_{32}\simeq\Delta m^2_{31}= 2.5 \times 10^{-3}\ \mathrm{eV}^2$. For the Earth-matter background, we adopt a constant mass density of $\rho = 5.51\ \mathrm{g/cm^3}$, corresponding to the mean density of the Earth, and assume a constant electron fraction $Y_e \simeq 0.5$. This simplified constant-density model is used as a controlled approximation for the LGI analysis presented below. Our primary aim here is to isolate and examine how matter effects modify the temporal correlations and the corresponding LGI parameter $K_4$, rather than to provide a realistic prediction for a specific neutrino trajectory through the Earth. A trajectory-dependent treatment would require the electron density to vary along the propagation path according to an adopted Earth density profile. We leave such a treatment, which is necessary for a detailed phenomenological comparison with atmospheric-neutrino experiments, beyond the scope of the present study.
	
	Evaluating the resonance condition of Eq.~\eqref{eq:Eres} numerically is most
	conveniently done using the standard relation
	$2\sqrt{2}\,G_F N_A E \simeq 1.52\times10^{-4}\ \mathrm{eV}^2\,
	(E/\mathrm{GeV})(Y_e \rho/\mathrm{g\,cm}^{-3})$
	\cite{Wolfenstein1978,Mikheev1985}, so that
	\begin{equation}
		E_{\text{res}} \simeq \frac{\Delta m^2 \cos 2\theta}{1.52\times10^{-4}\,
			Y_e\, \rho}\ \ \mathrm{GeV}. \label{eq:Eresnum}
	\end{equation}
	For the atmospheric-scale channel ($\Delta m_{32}^2=2.5\times10^{-3}\,\mathrm{eV}^2$, $\theta_{13}\simeq8.56^\circ$), this gives $E_{\text{res}}^{\text{atm}}\simeq5.72\,\mathrm{GeV}$. Substituting the solar mass splitting and $\theta_{12}$ gives the formal terrestrial-medium value $E_{\text{res}}^{(21)}\simeq0.07\,\mathrm{GeV}$. The latter is well above the endpoint of the standard solar-neutrino spectrum (about $18.8$ MeV, including the rare hep component) \cite{bahcall1998hep}; it is therefore used only to illustrate the response of the solar-splitting parameter sector in the constant-density model, not as a resonance accessible to solar-origin neutrinos. The near-resonance and off-resonance choices in Figs.~\ref{fig:baseline} and \ref{fig:energy} should be read with this distinction in mind.
	
	In the relativistic limit, the oscillation phase reduces to
	\begin{equation}
		\psi_{a;ij} \cong \psi_a = \frac{\Delta m_m^2}{4E} L_a.
		\label{eq:phaserel}
	\end{equation}
	Moreover, a maximal transition probability due to the MSW mechanism is possible provided
	that the oscillation phase satisfies the constraint
	\begin{equation}
		\psi_a = \frac{2k+1}{2}\pi, \hspace{1cm} k=0,1,2,\dots \label{eq:2}
	\end{equation}
	According to Eq.~\eqref{ncij}, the four two-time correlation functions entering
	$K_4$ are reconstructed from oscillation probabilities measured at different
	baselines $L_a$. Correspondingly, the sum rule of Eq.~\eqref{eq:sumrule},
	expressed in terms of the propagation lengths, becomes
	\begin{equation}
		L_a + L_b + L_c = L_d. \label{eq:sumruleL}
	\end{equation}
	Imposing the constraint given in Eq.~\eqref{eq:2} implies that a convenient
	choice of baselines satisfying this relation is
	\begin{equation}
		L_a = L, \quad L_b = 3L, \quad L_c = 5L, \quad L_d = 9L.
		\label{eq:baselines}
	\end{equation}
	Using these relations, together with the oscillation parameters introduced
	above, we compute the LGI parameter $K_4$ as a function of energy and baseline
	length, evaluated separately for the vacuum case ($V=0$) and the matter case
	($V \neq 0$).
	
	\subsubsection{Vacuum Case ($V=0$)}
	
\begin{figure}[htbp]
		\centering
		\includegraphics[width=0.8\textwidth]{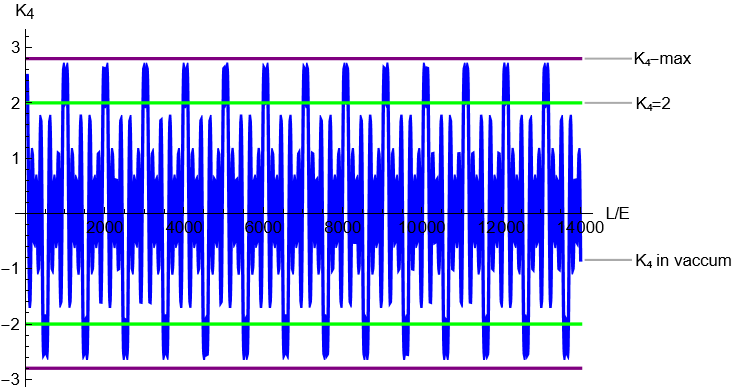}
		\caption{The LGI parameter $K_4$ in vacuum ($V=0$) as a function of the
			scaling variable $L/E$, for the baseline configuration given in
			Eq.~\eqref{eq:baselines}. The curve is plotted for atmospheric neutrinos
			undergoing $2 \to 3$ flavor transitions, with
			$\Delta m^2 = 2.45 \times 10^{-3}\,\mathrm{eV}^2$ and
			$\theta = 49.02^\circ$. The green horizontal lines mark the classical
			bounds $K_4 = \pm 2$, while the purple horizontal lines mark the extrema
			of the quantum prediction, $K_4^{\max} = \pm 2.8$; regions where the curve
			passes beyond the classical bounds indicate a violation of the LGI.}
		\label{fig:vacuum}
	\end{figure}

	In vacuum ($V=0$), the effective parameters reduce to their vacuum values,
	$\Delta m_m^2 \to \Delta m^2$ and $\theta_m \to \theta$. Substituting this
	limit and the baseline choice of Eq.~\eqref{eq:baselines} into the correlation
	function of Eq.~\eqref{ncij} yields
	\begin{align}
		K_{4} &= 2 - 2\Bigg( \sin^2\!\left[\frac{1.27\, L\, \Delta m^2}{E}\right]
		+ \sin^2\!\left[\frac{3.81\, L\, \Delta m^2}{E}\right]
		+ \sin^2\!\left[\frac{6.35\, L\, \Delta m^2}{E}\right] \nonumber \\
		&\quad - \sin^2\!\left[\frac{11.43\, L\, \Delta m^2}{E}\right] \Bigg)\sin^2 2\theta .
		\label{eq:KV4}
	\end{align}
	Because $K_4$ depends on energy and baseline only through the ratio $L/E$ at
	fixed oscillation parameters, it can be plotted as a function of this single
	variable, allowing regions that violate the full classical bound, $|K_4|>2$,
	to be identified directly. Figure~\ref{fig:vacuum} shows $K_4$ as a function of
	$L/E$ in vacuum for the baseline configuration of Eq.~\eqref{eq:baselines},
	using the atmospheric neutrino parameters $\Delta m^2 = 2.45 \times 10^{-3}\,
	\mathrm{eV}^2$ and $\theta = 49.02^\circ$ for the 2--3 sector. The vacuum
	result serves as a benchmark against which matter effects are compared in the
	following subsection.
	
	\subsubsection{Matter Case ($V \neq 0$)}
	In the presence of matter ($V \neq 0$), the effective mass-squared difference
	$\Delta m_m^2$ and mixing angle $\theta_m$, given in
	Eqs.~\eqref{eq:tan2thetam} and \eqref{eq:dm2m}, replace their vacuum
	counterparts in Eq.~\eqref{eq:KV4}. The resulting matter-modified expression is
	\begin{align}
		K_{4} &= 2 - 2\Bigg( \sin^2\!\left[\frac{1.27\, L\, \Delta m_m^2}{E}\right]
		+ \sin^2\!\left[\frac{3.81\, L\, \Delta m_m^2}{E}\right]
		+ \sin^2\!\left[\frac{6.35\, L\, \Delta m_m^2}{E}\right] \nonumber \\
		&\quad - \sin^2\!\left[\frac{11.43\, L\, \Delta m_m^2}{E}\right] \Bigg)\sin^2 2\theta_m.
		\label{eq:KM4}
	\end{align}
	The numerical coefficients in the arguments of the sine functions are unchanged
	from the vacuum case; only the matter-modified mass splitting $\Delta m_m^2$
	and the matter-enhanced mixing angle $\theta_m$ enter, controlling respectively
	the oscillation phases and the overall amplitude of $K_4$.
	
\begin{figure}[htbp]
		\centering
		\begin{subfigure}[b]{0.49\textwidth}
			\centering
			\includegraphics[width=\textwidth]{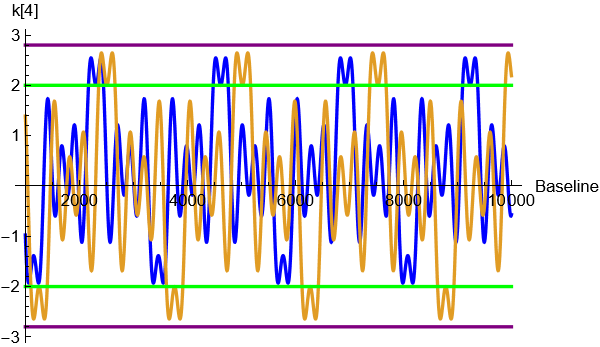}
			\caption{Near resonance, $E_{\text{res}} \simeq 0.07~\mathrm{GeV}$.}
			\label{fig:solar_near}
		\end{subfigure}
		\hfill
		\begin{subfigure}[b]{0.49\textwidth}
			\centering
			\includegraphics[width=\textwidth]{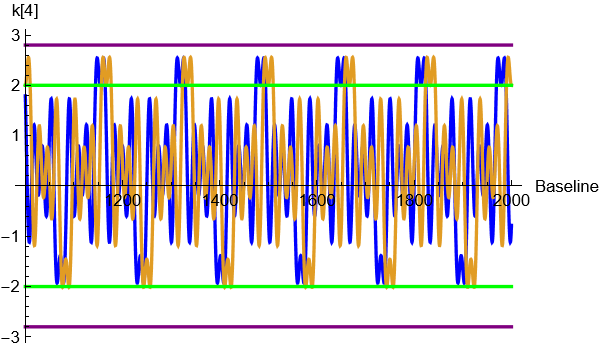}
			\caption{Far from resonance, $E = 0.005~\mathrm{GeV}$.}
			\label{fig:solar_far}
		\end{subfigure}
		
		\vspace{0.5em}
		\begin{subfigure}[b]{0.49\textwidth}
			\centering
			\includegraphics[width=\textwidth]{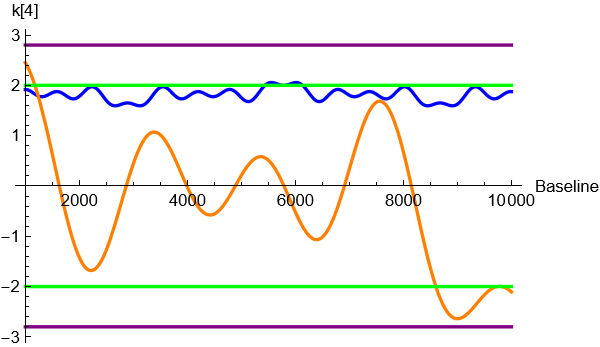}
			\caption{Near resonance, $E_{\text{res}} \simeq 5.73~\mathrm{GeV}$.}
			\label{fig:atm_near}
		\end{subfigure}
		\hfill
		\begin{subfigure}[b]{0.49\textwidth}
			\centering
			\includegraphics[width=\textwidth]{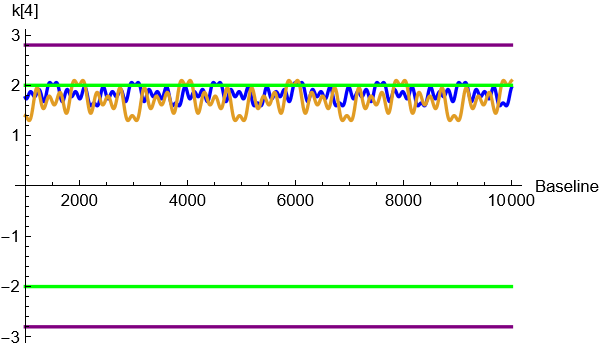}
			\caption{Away from resonance, $E = 1.5~\mathrm{GeV}$.}
			\label{fig:atm_far}
		\end{subfigure}
		
		\caption{The LGI parameter $K_4$ versus baseline length $L$ for the solar-splitting parameter sector (top row) and the atmospheric-scale sector (bottom row), using the baseline pattern in Eq.~\eqref{eq:baselines}. Blue denotes vacuum and orange denotes constant-density matter. The top-left panel uses the formal $1$--$2$ terrestrial-matter resonance energy, $0.07$ GeV, which is not accessible to solar-origin neutrinos; it is shown only to compare the parameter sector in the idealized model. The off-resonance panels show that the matter--vacuum difference is smaller for the selected parameters, although residual phase-dependent differences remain. Green lines mark the classical bounds $K_4=\pm2$, and purple lines indicate the quoted quantum extrema $K_4^{\max}=\pm2.8$.}
		\label{fig:baseline}
	\end{figure}

	To illustrate the effect of the MSW resonance on the flavor-based LGI quantity, we examine the baseline dependence of $K_4$ at selected energies. Figure~\ref{fig:baseline} shows the solar-splitting parameter sector (top row) and the atmospheric-scale sector (bottom row). The blue curve is the vacuum result and the orange curve is the constant-density matter result. In the selected near-resonance examples, the matter curve departs markedly from the vacuum curve, most conspicuously in the atmospheric-scale panel. In the selected off-resonance examples, the difference is generally smaller, although residual phase-dependent differences remain, especially in the solar-splitting panel. 
	
	The enhanced violation near resonance therefore confirms that the MSW
	mechanism amplifies the temporal quantum correlations responsible for LGI
	violation, and the resonance region provides the optimal conditions for
	observing a large departure from the classical bound.
	
\begin{figure}[htbp]
		\centering
		\includegraphics[width=0.49\textwidth]{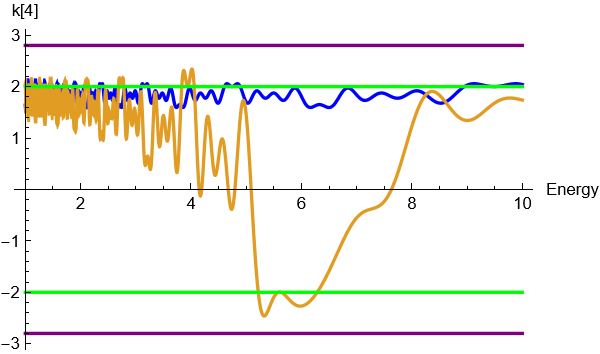}
		\hfill
		\includegraphics[width=0.49\textwidth]{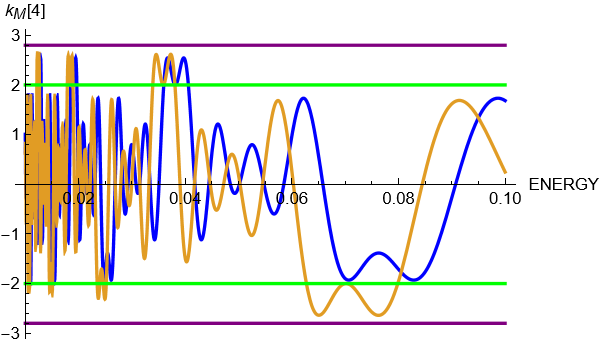}
		\caption{The LGI parameter $K_4$ as a function of neutrino energy $E$ at
			fixed baseline lengths, for the baseline configuration given in
			Eq.~\eqref{eq:baselines}. (Left panel) Atmospheric neutrinos with
			$L = 9580~\mathrm{km}$. (Right panel) Solar neutrinos with
			$L = 1258~\mathrm{km}$. In both panels, the blue curve shows the vacuum
			result ($V=0$), the other curve shows the matter result ($V \neq 0$), and
			the vertical dashed line marks the corresponding resonance energy. The
			green horizontal lines mark the classical bounds $K_4 = \pm 2$, while the
			purple horizontal lines mark the extrema of the quantum prediction,
			$K_4^{\max} = \pm 2.8$.}
		\label{fig:energy}
	\end{figure}

	The energy dependence is shown in Fig.~\ref{fig:energy}. The left panel uses the illustrative baseline $L=9580$ km with atmospheric-scale parameters; the right panel uses $L=1258$ km with the solar-splitting parameter sector. 
	In both panels, the matter curve departs most strongly from the vacuum curve in the vicinity of the resonance for the parameter choices shown. Away from resonance the curves can still differ because the effective phase is energy dependent. Consequently, the plotted energy dependence is oscillatory rather than a universal narrow peak. The $0.07$ GeV feature in the right panel is a formal matter response of the $1$--$2$ parameter sector, not a prediction for solar-origin neutrinos at that energy.
	
	The numerical results presented above establish a clear and physically
	meaningful connection between the MSW resonance and the violation of the LGI.
	The enhancement of the effective mixing angle at resonance maximizes the
	amplitude of the oscillatory terms in Eq.~\eqref{eq:KM4}, while the energy
	dependence of $\Delta m_m^2$ tunes the oscillation phases to optimize the sum
	of the four correlation functions. Far from resonance, the matter effects
	become negligible and the system reverts to the vacuum behavior shown by the
	blue curves in Figs.~\ref{fig:baseline} and \ref{fig:energy}, and by
	Fig.~\ref{fig:vacuum}. These findings demonstrate that Earth's matter effects
	not only influence flavor conversion probabilities but also provide a
	controllable mechanism for amplifying temporal quantum correlations, making
	atmospheric neutrinos ideal probes of macroscopic realism.
	
	\section{Conclusion}\label{sec:conclusion}
	
	We have studied the four-time LGI, \(K_4\), for two-flavor neutrino oscillations using the PDM formalism. Within this framework, the Horodecki theorem \cite{Horodecki1995} can be adapted to the PDM. For any two-time Pauli-correlation tensor, unitary qubit evolution induces an orthogonal rotation, leading to \(M_\rho=2\) and hence \(K_4^{\max}=2\sqrt{2}\) in the optimized temporal construction. This is, of course, a statement about the existence of suitable measurement settings within that construction; it does not imply that the fixed flavor observable \(Q=\sigma_z\) violates the LGI at every energy, baseline, or matter potential. Accordingly, the flavor-based \(K_4\) is evaluated explicitly in the numerical examples.
	
	Our constant-density analysis shows that matter effects can produce a pronounced difference between the vacuum and matter predictions near the MSW resonance for the selected parameters, whereas the selected off-resonance examples generally exhibit a smaller difference. The magnitude and sign of the departure still depend on the oscillation phase, so the resonance does not guarantee a violation for arbitrary baselines. We emphasize that the resonance at about $0.07$ GeV, obtained from the solar mass-splitting parameters in terrestrial matter, is used only as a formal parameter-sector illustration and is not attributed to solar-origin neutrinos, whose spectrum ends below about $18.8$ MeV \cite{bahcall1998hep}.
	
	These calculations are intended to isolate how the matter-modified mixing angle and phase influence an ideal two-flavor \(K_4\), not to provide a realistic Earth propagation calculation or an estimate of detectable violation events. Future work should incorporate realistic Earth density profiles such as PREM \cite{dziewonski1981preliminary}, extend the calculation to three flavors and to neutrinos/antineutrinos, account for energy and angular resolution and backgrounds, and perform a statistically calibrated analysis. Atmospheric-neutrino data and facilities such as DUNE \cite{acciarri2016long} and Hyper-Kamiokande \cite{abe2018hyper} provide strong motivation for such a dedicated future study, but no experimental sensitivity claim is made here. More broadly, our results establish that while the MSW effect can act as a controllable amplifier of temporal quantum correlations, the observation of LGI violation in neutrino oscillations requires both a careful choice of measurement settings and an appropriate baseline configuration.
	
	\section*{Acknowledgements}
	The authors would like to thank the research Deputy of the University of Qom for support.
	

\end{document}